\documentclass[sn-basic]{sn-jnl}
\jyear{2022}%

\theoremstyle{thmstyleone}%
\usepackage{makecell}
\usepackage{booktabs}

\theoremstyle{thmstyletwo}%

\usepackage{float}
\usepackage{rotating}

 \usepackage{csquotes}
 \usepackage{lipsum}
 
\usepackage{hyperref}
\usepackage{academicons}
\usepackage{xcolor}

\newcommand{\orcid}[1]{\href{https://orcid.org/#1}{\textcolor[HTML]{A6CE39}{\aiOrcid}}}

\theoremstyle{thmstylethree}%

\usepackage{physics}
\usepackage{subfig}
\usepackage{epigraph}

\begin{document}

\title{The influence of color on prices of abstract paintings}


\author[1]{\fnm{Maksim} \sur{Borisov} } \email{maksim.borisov.2013@gmail.com}  

\author[1]{\fnm{Valeria} \sur{Kolycheva }}\email{v.kolycheva@spbu.ru}

\author[2]{\fnm{Alexander} \sur{Semenov}}\email{asemenov@ufl.edu}

\author[1]{\fnm{Dmitry} \sur{Grigoriev}}\email{d.a.grigoriev@spbu.ru}

\affil*[1]{\orgdiv{Center of Econometrics and Business Analytics (CEBA), St. Petersburg State University}, \orgaddress{\street{ 7/9 Universitetskaya nab.}, \city{ St. Petersburg}, \postcode{199034}, \country{Russia}}}


\affil*[2]{\orgdiv{Herbert Wertheim College of Engineering}, \orgname{University of Florida}, \orgaddress{\street{1949 Stadium Rd}, \city{Gainesville}, \postcode{32611}, \state{FL}, \country{USA}}}


\abstract{Determination of price of an artwork is a fundamental problem in cultural economics. In this work we investigate what impact visual characteristics of a painting have on its price. We construct a number of visual features measuring complexity of the painting, its points of interest, segmentation-based features, local color features, and features based on Itten and Kandinsky theories, and utilize mixed-effects model to study impact of these features on the painting price. We analyze the influence of the color on the example of the most complex art style - abstractionism, by created Kandinsky, for which the color is the primary basis. We use Itten's theory - the most recognized color theory in art history, from which the largest number of subtheories was born. For this day it is taken as the base for teaching artists. We utilize novel dataset of 3,885 paintings collected from Christie's and Sotheby's and find that color harmony has some explanatory power, color complexity metrics are insignificant and color diversity explains price well.}

\keywords{paintings, abstractionism, color theory, Itten's color wheel, pricing, mixed-effects}

\pacs[JEL Classification]{Z11, C81}


\maketitle

\epigraph{\itshape  Color is a means of exercising direct influence upon the soul. color is the keyboard. The eye is the hammer, while the soul is a piano of many strings. The artist is the hand through which the medium of different keys causes the human soul to vibrate. }{---Wasily Wasilyevich Kandinsky (1866-1944), \textit{the founder of abstractionism}}

\epigraph{\itshape  We start with color.
}{---Mark Rothko (1903-1970), \textit{one of the most expensive abstract painters}}

\section{Introduction}\label{sec1}

The reaction  of people to a particular color can be determined by the cultural environment (in different cultures the same colors can symbolize different concepts) and the personal preferences of a particular person. For example, red in France is the color of aristocrats, as it is known. But the French led the Egyptians to complete bewilderment, as in Egypt, red is a symbol of mourning. Because of this, the Egyptians had the persistent impression that all French nobility was in an incessant state of mourning. In China, red is worn by brides as a symbol of endurance and faith. In Japan, red means danger, as well as anger and rage. A curious fact is Americans will never agree with the French, because green in America is security, and in France it symbolizes crime \citep{adams1973cross}.

One of the most profound thinkers of color symbolism, Johann Wolfgang von Goethe, believed that color, regardless of the structure and shape of the material to which it belongs, has a certain effect on a person’s mental mood~\cite{text1}. Goethe writes about the perception of color and the influence it has on consciousness. From his point of view, there are special colors that can cause special conditions or correct existing ones. He offers various forms of working with color: simply looking at it, surround yourself with objects of a certain color, use different types of colored glasses.

Goethe's research was continued and developed by Johannes Itten (1888-1967), the creator of the most famous color sphere \cite{itten1970elements}. In his opinion, the fact of the influence of colors on people's emotional well-being is obvious. He argued that we treat every shade of color with a subjective assessment, since inevitably the human soul responds to every shade and tone of color with a feeling of either sympathy or antipathy. The characteristic of color and its effect on a person is determined by its location in relation to other accompanying colors.

Abstractionism, abstract art (Latin \textit{abstractio} for ``removal, distraction'') is an art style that refused to depict forms that are close to reality. One of the goals of abstractionism is to achieve harmonization by depicting certain color combinations, causing the viewer to feel the completeness of the composition. In 1910-1915, painters in Russia and Western Europe began to create abstract art works. The year of birth of non-objective art is considered to be 1910, when in Murnau, Germany, Kandinsky created his first abstract composition. The aesthetic concepts of the first abstractionists assumed that artistic creativity reflects the laws of the universe, hidden behind external, superficial phenomena of reality. These patterns, intuitively comprehended by the artist, are expressed in the artwork through the ratio of abstract forms (color spots and lines).

In abstract painting, color is one of the main visual means, in some types of abstraction it is the most important, and sometimes just the only one. In an abstract painting, painting in its classical sense is entirely absent. The colors of paints do not serve to depict the surrounding world, but to express visually the thoughts, feelings, impressions and experiences of an abstractionist, received by him from observation or even from a collision with the phenomena of the world around him. The only thing that can be said with certainty is that the role of color is one of the most important in the transmission of inner feelings and sensations.

The most important and valuable property of color in abstract art is the ability of different colors and their shades to evoke psycho-physiological feelings and sensations in people. The viewer feels these color emotions without even guessing why they feel them. There is an almost complete analogy of color with musical sounds that create their own emotions in listeners of good music.

The philosophical basis for the depiction of non-objective reality were the ideas of the French philosopher Henri-Louis Bergson and the German philosopher Friedrich Wilhelm Ostwald. Bergson, in his doctrine of the ``vital impulse'' (élan vital), arguing that the essence of life can only be realized with the help of intuition, since life is perceived directly, i.e. through one's experiences. Indeed, this is the only way to perceive abstract images according to the philosopher \cite{text_4}. The intelligence is powerless in this case, since it operates with material, spatial objects.

Ostwald, the founder of ``energetic theory'' (the philosophical doctrine of ``energetics''), believed that energy is primary in relation to matter~\cite{text_5}. Such a worldview formed the idea of the existence of the spiritual energy of the cosmos, which gives rise to order from chaos. Abstract canvases are practically permeated with such energy and carry a strong emotional message for every viewer. These philosophical ideas determined the main pictorial motif of abstractionism~--- the space of dematerialized phenomena, i.e. devoid of material substrate.

The effect of color on a person was considered by Kandinsky in his work ``On the Spiritual in Art''. The effect of color on a person occurs in two ways: ``the first is a purely physical effect when the eye itself is enchanted by beauty and the multiple delight of color. The observer is pleased. He experiences a pleasure similar to that enjoyed by an epicure in tasting a delicacy. The eye is stimulated as the tongue is titillated by a spicy dish. Or it is refreshed and soothed as a finger touching ice. ... The bright yellow of a lemon hurts the eye after a while, as a shrill trumpet note may disturb the ear. The eye becomes restless, is unable to fix its gaze for any length of time, and seeks distraction and rest in blue or green.... The eye is attracted by light colors and still more by the lightest, warmest ones.''\cite{text_6}.

Kandinsky noted that color has a physical and mental impact on a person. In addition, some paints seem soft, and others seem hard: ``In the case of the eye, some colors can look sharp or piercing, while others appear smooth like velvet, so that one feels inclined to stroke them (dark ultra-marine, chrome oxide green and rose madder); even the distinction between the warmth and coldness of a shade is based on such feelings. Some colors appear soft (rose madder) and others seem to be cold and hard (cobalt green, blue-green oxide), so that even when such color is freshly squeezed from a tube, it has a dry appearance''\cite{text_6}. 


\begin{displayquote}








Kandinsky also wrote about the Itten sphere, describing the moods and associations that each of the sections could evoke in a viewer. He ascribes ``deep seriousness'' to blue, ``life and growth'' to green, ``pure joy and infinite purity'' to white, and vice versa ``deepest sorrow and death'' to black. The colour obtained by mixing white and black, grey, Kandinsky believed to be ``desolation''. Red, in Kandinsky's opinion, conveys ``strength, energy, ambition'', brown~--- ``inner beauty'', orange~--- ``emanating happiness and health'', and finally, violet is ``frailty, expiring sadness''.
\cite{text_6}
\end{displayquote}

Blue certainly became the main color of abstractionists. In nature, blue is the color of water and air space, the main sources of life. In the history of art, blue is traditionally considered as a symbol of human spiritual aspirations. On the eve of the First World War, Wassily Kandinsky and Franz Marc (1880-1916) enthusiastically hoped for the coming ``Epoch of Great Spiritual'', which would unite all kinds of art and culture. The bearer of such a universal meaning for them was ``The Blue Rider'' depicted on the cover of the almanac (1911), which gave the name to the group. Mark developed his own theory of color, where he gave each of the main colors a special spiritual meaning: blue embodied for him an ``ascetic'' beginning. 
In today's world, blue is one of the most frequently chosen colors as favorite or preferred in some sorts of polls. \cite{singh2006impact}. Blue is the official color of international organizations, including UN, CoE, UNESCO, EU and NATO. Blue is the most common color of military uniforms and men’s business suits. The largest social networks~--- Facebook, Twitter, LinkedIn, VK, LiveJournal, Tumblr, Foursquare and many others~--- have chosen blue for their design.

Thus, in abstract compositions, color serves as the main means for conveying emotions and moods. With its help, artists can give to their pictures a certain emotional expressiveness, creating different moods. ``These feelings quoted as parallels to these colors (such as joy, sorrow) express the material conditions of the soul. Variations of color, like those of music, are of a much subtler nature, and awaken in the soul much finer vibrations than words could''. \cite{text_6} 

\textcolor{black}{In the scope of this article we will answer a research question: How do color metrics, such as color harmony according to Itten, color complexity and others, influence the abstract painting price?}


We construct a number of visual
features measuring complexity of the painting, its points of interest,
segmentation-based features, local color features, and six color harmony features based
on Itten and Kandinsky theories, and utilize mixed-effects model to
study impact of these features on the painting price. We analyze the
influence of the color on the example of the most complex art style -
abstractionism, by created Kandinsky, for which the color is the primary basis. We use Itten’s theory - the most recognized color theory
in art history, from which the largest number of subtheories was born.
For this day it is taken as the base for teaching artists. We utilize
novel dataset of 3,885 paintings collected from Christie’s and Sotheby’s
and find that color harmony has some explanatory power, color complexity metrics are insignificant and color diversity explains price well.

Our study makes several contributions to the current literature.
\begin{enumerate}

\item We apply Itten color theory to price-harmony relationship.
\item We study the effect of visual complexity of the paintings on their price. Visual complexity is operationalized as a number of features constructed from the digital image content.
\item We utilize the mixed-effects model to take into account the effect of the author, visual features, and other characteristics.
    
\end{enumerate}

The paper is structured as follows: Section~\ref{sec2} contains a survey of related literature, Section~\ref{sec3} describes our dataset, and Section~\ref{sec4} contains the description of the statistical model. Section~\ref{sec5} discusses digital imaging background and introduces constructed features, the next Section~\ref{sec6} presents empirical results, and the final Section~\ref{sec7} concludes the paper.

\section{Related Work}\label{sec2}
In art literature, the relationship between the price of a painting and
its attributes such as size~\cite{schonfeld2007effects}, material and signature, sale conditions such
as year, salesroom and sale location has been studied with the hedonic model since the 70's (e.g., \cite{buelens1993revisiting}, \cite{worthington2004art}, \cite{chanel1995art}). The painting size is normally found to be a significant explanatory variable for price.

In the past years, quantitative studies of color in art have been conducted multiple times (such as \cite{romero2018computational}, \cite{montagner2018supporting}, and \cite{farrell2021gender}). Their work suggest that by analysing art one can obtain results useful for understanding human perception. \cite{graham2010statistical}.

The paper~\cite{guo2018assessment} proposed 29 global, local, and salient region features
which represent distribution of compositions, colors, and contents. The results show that a regression model with these features has a good ability of measuring aesthetic quality, beauty, and
the people's liking of the color composition of paintings \cite{guo2018assessment}.

The connection between color composition of an artwork and its price is still a new research area. Article \cite{graham2010preference} concludes that the selling price is not predictive of preference, while shared preferences may to some extent be predictable based on image statistics \cite{graham2010preference}. Paper~\cite{stepanova2019impact} finds significant evidence that contrastive paintings (those with a high diversity of colors) carry premium price than equivalent monochrome artworks.
Empirical results presented in \cite{pownall2016pricing} present significant evidence of intense colors fetching a premium
over equivalent artworks which are less intense in color. Furthermore, darkness carries a premium over lightness.

Previous studies of harmony
of color combinations have produced confusing results. For
example, some claim that harmony increases with hue
similarity, whereas others claim that it decreases \cite{schloss2011aesthetic}.

By color harmony we mean the craft of developing themes from systematic color relationships capable of serving as a basis of a composition. Johannes Itten suggests using the color wheel to construct harmonic color combinations \cite{itten1970elements}.

Ventura Charlin and Arturo Cifuentes proposed a general framework to study the price-color relationship in paintings. They have found  that the dominant colors and the diversity of the color palette, are
by far the most relevant attributes that influence the price; color harmony and
color emotions hold almost no explanatory power during 2006-2018s auction data of Mark Rothko's post-1950 paintings \cite{charlin2021general}.

Although some research has been carried out on price-color relationship, no single study exists which adequately covers effect of painting visual complexity on its price. 
Ventura  Charlin  and  Arturo  Cifuentes have found insignificance of color harmony. However, these findings are limited by the use of recent model by Ou and Luo to describe color harmony. 
From our knowledge, previous studies of art pricing have not dealt with multiple authors at once.
Therefore, one does not take advantage of the information in data from other authors. This can also make the results noisy, since the estimates from each model are not based on very much data.

\section{Data}\label{sec3}
We analyze 3885 paintings collected from Sotheby's and Christies auction houses. The main style of paintings is abstract. Our dataset includes paintings from such authors as Franz Kline, Mark Rothko, Cy Twombly, Alexander Calder, Joan Miró, Joan Mitchell, Lucio Fontana, Philip Guston, Richard Diebenkorn, Wassily Kandinsky, Willem de Kooning, Zao Wou-ki, Henry Moore.
The minimum number of paintings of a single author is 114, and the maximum is 655. The selected images belong to high price category, with a mean normalized price totaling around 11.9 million dollars. 
\textcolor{black}{Author birth years range is from 1866 to 1928, and the most frequent material is paper.
Ninety percent of paintings have a signature, and about half of all paintings were obtained from Sotheby's. The most used color in the paintings is green.}
Descriptive statistics are presented in \hyperref[Table 1]{Table 1} and \hyperref[Table 3]{Table 3}. Description of authors is presented in \hyperref[Table 2]{Table 2}.

\begin{table}[h!]\label{Table 1}
    \centering
    \caption{Descriptive statistics grouped by authors.}
    \scalebox{1}{

      \begin{tabular}{lrllrrr}
\toprule
            \thead{Author} &  \thead{Number\\of\\ paintings}  & \thead{Min-Max \\price\\(M $\$$)} &  \thead{Mean\\price \\(M $\$$)}  &  Christie's &  Sotheby's \\
\midrule

       Mark Rothko &                  114  &          9.6 - 18.4 &        14.9 &        63.0 &       51.0 \\
     Philip Guston &                  144  &          8.2 - 17.2 &        12.4 &        83.0 &       61.0 \\
Richard Diebenkorn &                  167 &          6.5 - 17.0 &        12.9 &       111.0 &       56.0 \\

     Lucio Fontana &                  168  &          6.9 - 16.9 &        12.6 &       110.0 &       58.0 \\

       Franz Kline &                  170  &          8.5 - 17.6 &        12.2 &        90.0 &       80.0 \\
       Joan Mitchell &                  188  &          7.0 - 16.6 &        13.2 &        98.0 &       90.0 \\

       Cy Twombly &                  294  &          6.6 - 18.1 &        12.6 &       155.0 &      139.0 \\
       Zao Wou-ki &                  305  &          6.9 - 18.0 &        12.1 &        75.0 &      230.0 \\

 Wassily Kandinsky &                  333  &          7.1 - 17.4 &        12.3 &       205.0 &      128.0 \\

       Henry Moore &                  363  &          5.7 - 14.9 &         9.5 &       190.0 &      173.0 \\
        Willem de Kooning &                  402 &          6.9 - 18.1 &        12.5 &       235.0 &      167.0 \\

         Joan Miró &                  582  &          6.4 - 17.0 &        11.6 &       197.0 &      385.0 \\

  Alexander Calder &                  655  &          7.5 - 15.1 &        11.0 &       279.0 &      376.0 \\
  
\bottomrule
\end{tabular}
}
\end{table}

\begin{table}\label{Table 2}
    \centering
    \caption{Description of authors.}
      \begin{tabular}{lrll}
\toprule
            \thead{Author} &  \thead{Years\\of\\ life} & \thead{Nationality} & \thead{Description} \\
\midrule

 \thead{Wassily\\Kandinsky} &                  \thead{1866–\\1944}  &          \thead{Russian-\\German} &       \thead{Founder and theorist of abstractionism.} \\

 \thead{Joan\\Miró} &                  \thead{1893–\\1983}  &           \thead{Catalan/\\Spanish} &      \thead{Notable for his interest in the unconscious or \\ the subconscious mind, reflected in \\  his re-creation of the childlike.} \\

  \thead{Alexander\\Calder} &                  \thead{1898-\\1976}  &         \thead{American} &   \thead{His paintings foreshadowed first\\ abstract sculptures with real movement.} \\
  
      \thead{Henry\\Moore} &                  \thead{1898–\\1986}  &         \thead{English} &       \thead{Associated with subtle abstraction inspired by \\African and other non-Western art,\\ combined with bold shapes.} \\
       
            \thead{Lucio\\Fontana} &                  \thead{1899–\\1968}  &         \thead{Argentine-\\Italian} &     \thead{The most characteristic works \\  for him were paintings with slits and gaps.}\\

               \thead{Mark\\Rothko} &                 \thead{1903–\\1970} &         \thead{Russian-\\American} &      \thead{One of the creators of color field painting, \\ which is based on the use of large planes of uniform \\ colors, close in tone and not centered.}\\

 \thead{Willem\\de\\Kooning} &                  \thead{1904–\\1997} &          \thead{Dutch-\\American} &       \thead{Followed the path of primitivization,\\ developed so-called "figurative abstractions"} \\

       \thead{Franz\\Kline}&                  \thead{1910–\\1962}  &         \thead{American} &        \thead{Associated with canvases with large-format, contrasting style.} \\
       
                \thead{Philip\\Guston} &                  \thead{1913–\\1980}  &        \thead{Canadian-\\American} &     \thead{ Creating new paintings, he often\\ used fragments of previous canvases.} \\

        \thead{Zao\\Wou-ki} &                  \thead{1920–\\2013}  &        \thead{Chinese-\\French} &        \thead{Developed his own non-figurative style, \\combining the achievements of Western abstractionism \\ with Chinese painting traditions.} \\

                \thead{Richard\\Diebenkorn} &                  \thead{1922–\\1993} &        \thead{American} &        \thead{Known for his extensive series of geometric, \\lyrical abstract paintings, named Ocean Park paintings.} \\

            \thead{Joan\\Mitchell} &                  \thead{1925–\\1992}  &          \thead{American-\\French} &   \thead{Known for her emotionally intense \\ style and gestural brushwork.}  \\

               \thead{Cy\\Twombly} &                  \thead{1928-\\2011}  &          \thead{American-\\Italian}  &      \thead{His best-known works are typically large-scale, \\ freely-scribbled, calligraphic and graffiti-like.} \\        
\bottomrule
\end{tabular}
\end{table}

 \begin{table}[h!]\label{Table 3}
  \caption{Descriptive statistics.}
 \begin{center}
     
\begin{tabular}{lrrrrrrrr}
\toprule
{}  &   mean &    std &    min &    25\% &    50\% &    75\% &     max \\
\midrule
normalized\_price      &   11.9 &    2.3 &    5.7 &   10.4 &   11.7 &   13.5 &    18.4 \\
square\_m              &    0.7 &    1.0 &    0.0 &    0.1 &    0.4 &    0.8 &    11.2 \\
ExhibitedNum          &    1.2 &    2.4 &    0.0 &    0.0 &    0.0 &    1.0 &    29.0 \\
ProvenanceNum         &    2.8 &    2.2 &    0.0 &    1.0 &    3.0 &    4.0 &    18.0 \\
LiteratureNum         &    0.9 &    1.8 &    0.0 &    0.0 &    0.0 &    1.0 &    30.0 \\
date\_of\_birth         & 1902.9 &   16.1 & 1866.0 & 1898.0 & 1899.0 & 1913.0 &  1928.0 \\
oil                   &    0.4 &    0.5 &    0.0 &    0.0 &    0.0 &    1.0 &     1.0 \\
ink                   &    0.2 &    0.4 &    0.0 &    0.0 &    0.0 &    0.0 &     1.0 \\
gouache               &    0.2 &    0.4 &    0.0 &    0.0 &    0.0 &    0.0 &     1.0 \\
lithograph            &    0.1 &    0.3 &    0.0 &    0.0 &    0.0 &    0.0 &     1.0 \\
canvas                &    0.2 &    0.4 &    0.0 &    0.0 &    0.0 &    0.0 &     1.0 \\
paper                 &    0.7 &    0.5 &    0.0 &    0.0 &    1.0 &    1.0 &     1.0 \\
Christies             &    0.5 &    0.5 &    0.0 &    0.0 &    0.0 &    1.0 &     1.0 \\
Sothebys              &    0.5 &    0.5 &    0.0 &    0.0 &    1.0 &    1.0 &     1.0 \\
Sign                  &    0.9 &    0.3 &    0.0 &    1.0 &    1.0 &    1.0 &     1.0 \\
lines\_variance        &    0.1 &    0.1 &    0.0 &    0.0 &    0.1 &    0.1 &     0.4 \\
X\_contrst\_triad       &    0.1 &    0.3 &    0.0 &    0.0 &    0.0 &    0.0 &     1.0 \\
X\_classic\_triad       &    0.0 &    0.2 &    0.0 &    0.0 &    0.0 &    0.0 &     1.0 \\
X\_rectangle           &    0.0 &    0.1 &    0.0 &    0.0 &    0.0 &    0.0 &     1.0 \\
X\_analog\_triad        &    0.4 &    0.5 &    0.0 &    0.0 &    0.0 &    0.9 &     1.0 \\
X\_quad                &    0.0 &    0.1 &    0.0 &    0.0 &    0.0 &    0.0 &     1.0 \\
X\_comp                &    0.2 &    0.4 &    0.0 &    0.0 &    0.0 &    0.0 &     1.0 \\
ccm                   &    1.0 &    0.3 &    0.4 &    0.8 &    1.0 &    1.2 &     3.3 \\
points\_of\_interest    &   24.3 &   20.6 &    0.0 &    0.0 &   32.5 &   42.0 &    82.0 \\
fls\_h                 &   42.2 &   35.3 &   -1.0 &   20.2 &   26.0 &   51.5 &   176.4 \\
fls\_s                 &   66.8 &   62.1 &   -1.0 &   23.9 &   43.8 &   87.6 &   254.6 \\
fls\_v                 &  171.7 &   70.1 &   -1.0 &  119.4 &  200.9 &  229.2 &   255.0 \\
sls\_h                 &   43.2 &   35.9 &   -1.0 &   20.5 &   27.5 &   55.9 &   176.9 \\
sls\_s                 &   76.4 &   64.0 &   -1.0 &   29.3 &   55.8 &  104.4 &   254.7 \\
sls\_v                 &  163.7 &   61.8 &   -1.0 &  124.3 &  176.9 &  213.8 &   255.0 \\
contrast\_h            &   65.0 &   45.4 &   -1.0 &   22.8 &   63.6 &   93.8 &   168.9 \\
contrast\_s            &  112.8 &   70.8 &   -1.0 &   47.5 &  111.4 &  172.5 &   250.5 \\
contrast\_v            &  133.2 &   50.9 &   -1.0 &   95.7 &  137.5 &  175.4 &   236.8 \\
area\_of\_fls           &    0.5 &    0.3 &   -1.0 &    0.3 &    0.5 &    0.7 &     1.0 \\
area\_of\_sls           &    0.1 &    0.1 &   -1.0 &    0.1 &    0.1 &    0.2 &     0.5 \\
number\_of\_segments    &   38.2 &   31.1 &   -1.0 &   16.0 &   30.0 &   51.0 &   241.0 \\
shape\_complexity\_fls  &  309.1 & 2263.9 &   -1.0 &   17.3 &   45.9 &  109.4 & 56533.7 \\
shape\_complexity\_sls  & 1873.5 & 5770.4 &   -1.0 &   19.7 &   55.9 &  369.2 & 98055.8 \\
black                &    0.2 &    0.3 &    0.0 &    0.0 &    0.1 &    0.3 &    12.8 \\
CCT                   & 6238.1 & 1146.8 & 4043.8 & 5730.1 & 6085.9 & 6418.1 & 19054.5 \\
blue\_cluster          &    0.0 &    0.0 &    0.0 &    0.0 &    0.0 &    0.0 &     0.3 \\
green\_cluster         &    0.3 &    0.1 &    0.0 &    0.2 &    0.3 &    0.4 &     0.5 \\
red\_cluster           &    0.0 &    0.0 &    0.0 &    0.0 &    0.0 &    0.0 &     0.3 \\
yellow\_cluster        &    0.0 &    0.1 &    0.0 &    0.0 &    0.0 &    0.0 &     0.3 \\
\bottomrule
\end{tabular}
 \end{center}
\end{table}

\section{Model}\label{sec4}
Majority of studies use the hedonic model to find price determinants \cite{stepanova2019impact, cinefra2019determinants,pownall2016pricing}. In our case, we have multiple authors, which implies a different price range for each of them; this may result in heteroscedasticity, making the hedonic model unsuitable.
In order to account for the heteroscedasticity, we employ the linear mixed model: 

\begin{equation}
    y = X \beta + Zu + \varepsilon
\end{equation}
 $N$ is a number of observations. $y$ is a $N \times 1$ column vector, the log of painting price. X is $N \times 42$ matrix of the $42$ predictor variables. $\beta$ is $42 \times 1$ column vector of the fixed effects regression coefficients. $Z$ is $N \times J$ design matrix for the $1$ random effect and $J$ groups. $u$ is a $J \times 1$ vector of one random effect for $J$ groups. $\epsilon$ is $N \times 1$ column vector of the residuals. \\ 
In our case $N = 3885$, \verb+author+ is the grouping variable, and the number of authors is $J = 21$. We will focus on the basic mixed model with fixed effects for the features and a random intercept for each distinct value of group.

The regression coefficients estimated in R via the \verb+lme4+ package~\cite{bates2007lme4} represent the buyer’s willingness to pay a premium for a particular characteristic. The confidence intervals were build by computing a likelihood profile and finding the appropriate cutoffs based on the likelihood ratio test.

Suppose there are two sets of parameters, $\lambda$ and $\beta$, where $\lambda$ represents the vector of coefficients  associated with the random effect and $\beta$   represents regression coefficients. The likelihood $L(\lambda, \beta)$ is then a function
of $\lambda$ and $\beta$. We obtain the profile-likelihood function $L(\lambda)$ by replacing $\beta$ by its
maximum likelihood estimate $\widehat{\beta}(\lambda)$ at fixed values of $\lambda$.
\begin{equation}
    L(\lambda) = \max_{\beta}L(\lambda, \beta) = L(\lambda, \widehat{\beta}(\lambda))
\end{equation}
Specifically, likelihood profile method consists of two nested maximizations: $L(\lambda)$ is maximized
with respect to $\lambda$, where  $L(\lambda)$ is itself obtained by maximizing  $L(\lambda, \beta)$ with
respect to $\beta$.
\cite{murphy2000profile}

Marginal $R_{m}^{2}$ and conditional $R_{c}^{2}$ coefficient of determination were calculated as follows:

\begin{equation}
     R_{m}^{2} = \frac{\sigma^{2}_{f}}{\sigma^2_{f} + \sigma^{2}_{u} + \sigma^2_{\varepsilon}}
\end{equation}

\begin{equation}
     R_{c}^{2} = \frac{\sigma^{2}_{f} + \sigma^{2}_{u} }{\sigma^2_{f} + \sigma^{2}_{u} + \sigma^2_{\varepsilon}}
\end{equation}
Where $\sigma^{2}_{f} $ is the variance calculated from the fixed effect components (i.e. variance explained by
fixed factors). $\sigma^{2}_{u} $ is the variance component of the random
factor and $\sigma^2_{\varepsilon}$ is the residual variance.
Coefficients of determination are equal 0.672 and 0.773 respectively. \cite{nakagawa2013general}
\section{Price determinants}\label{sec5}

\subsection{Color description}
Paintings are represented by two-dimensional digital images.
Each image consists of pixels; if the height of an image is $N$ pixels, and the width is $M$ pixels, the image consists of $N \times M$ pixels, each is a point in color space. In our case, we employ the CIELab color space (abbreviated from Commission internationale de l'éclairage, the French title of the International Commission on Illumination)~\cite{zhang1996spatial}. It expresses color with three values $(L^*, a^*, b^*)$: $L^*$ represents perceptual lightness, $a^*$ and $b^*$ represent the four unique colors of human vision: red, green, blue, and yellow. CIELAB was designed as a perceptually uniform space, where a given numerical change corresponds to similar perceived change in color. While the CIELab space is not truly perceptually uniform, it nevertheless is useful in practical applications for detecting small differences in color. Alternative color spaces are RGB (red, green, blue), and CMYK (cyan, magenta, yellow).

In our work, we use visual features as independent variables in the mixed effects model~\cite{galecki2013linear}. These visual features represent measurable characteristics of the image, constructed from the raw pixels of the image. 
Price determinants can be divided into intrinsic characteristics and acquired characteristics. The intrinsic characteristics are the size, material, presence of signature and color distribution.
Acquired characteristics are number of exhibitions, the number of owners, the number of mentions in literature, the year its author was born, and the name of the auction house.

\subsection{Global color features}
\begin{figure}[h!]
    \centering
    \subfloat[\centering  Painting with White Lines, 1913,  \$41.60 million, \newline CCM = 1.24
    ]{{\includegraphics[width=4cm, height=5cm]{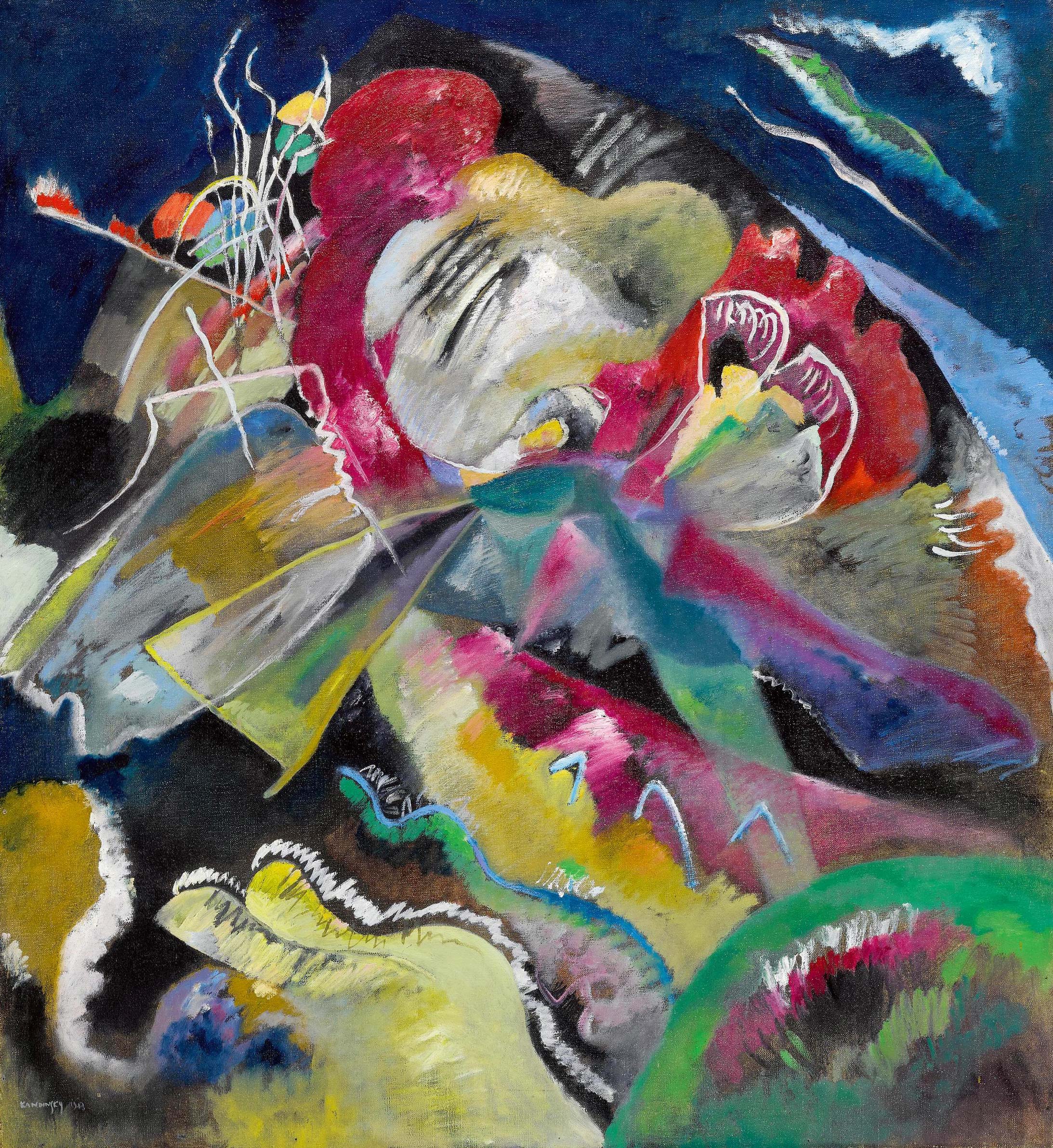}}}%
    \qquad
    \subfloat[\centering  Pink in Gray, 1926,     \$1.15 million,  CCM = 1.02]{{\includegraphics[ width=4cm, height=5cm]{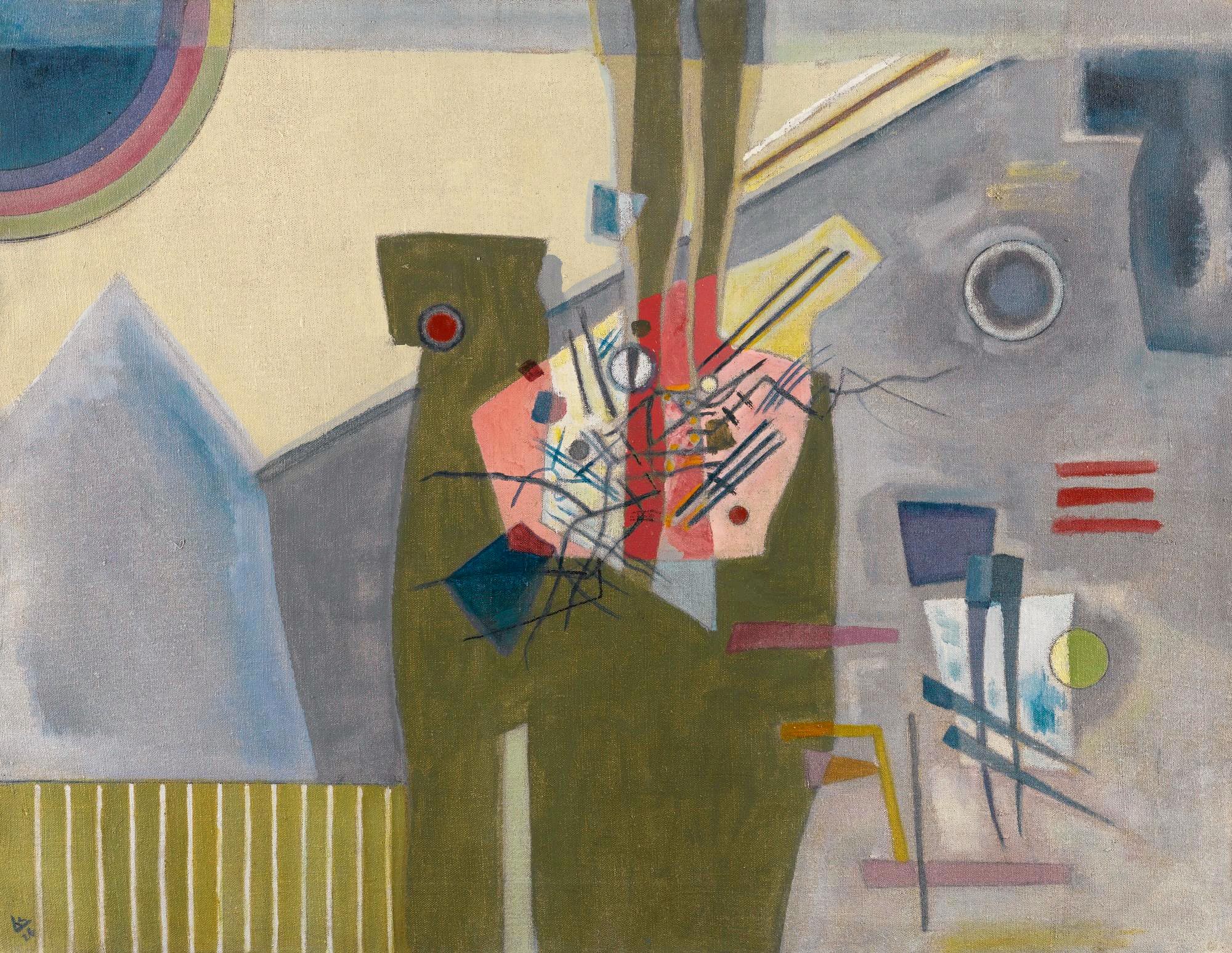}}}%
    \caption{\textcolor{black}{An expensive and an inexpensive artworks of Kandinsky. Painting on the left is more complex visually than the painting on the right, hence its color complexity measure is higher.} }%
    \label{fig:example_kandinsky}%
\end{figure}

Looking at a painting, people first obtain an impression of it before noticing subtler details~\cite{kim2013study}. Therefore,
global features may influence the first impression of human visual perception. Here, we will describe global features used in our mixed-effects model, namely color complexity measure, points of interest, edge density, tetrad, contrast triad, classic triad, complementary, rectangle, and analogue schemes. 
\textcolor{black}{We will illustrate our indexes using two works of Kandinsky, namely, ``Painting with White Lines'', 1913 and ``Pink in Gray'', 1926 with the corresponding prices of 41.60 and 1.15 millions U.S. Dollars respectively. The images are presented in Figure~\ref{fig:example_kandinsky}.}

\subsubsection{Color complexity measure}
To measure the complexity of the colors in a painting, we employ the color complexity measure (CCM)  ~\cite{ccm}. 

\textcolor{black}{Let  $(i, j)$ be the coordinates of some pixel in the image.
Consider a local window $\Omega_{(i, j)}$ with the center at  $(i, j)$. This window contains $5 \times 5$ pixels. The idea of CCM is to capture local color difference at the pixel with coordinates $(i,j)$ as follows:}
  \begin{equation}
      \psi(i,j) = \sum_{x, y \in \Omega_{(i, j)}} G_{\alpha}(\| c(x,y) - \bar{c}\|)
  \end{equation}

\textcolor{black}{in which $(x,y)$ is a pixel from a local window $\Omega_{(i, j)}$, $G_{\alpha}$ is the Gauss function, applied to additionally smooth results, $\| \cdot \|$ is some color difference measure (such as Euclidean distance), and $\bar{c}$ is the mean value of pixels from  $\Omega_{(i, j)}$}.

The color difference measure is important for a more accurate representation of the human visual perception of colors~\cite{guo2018assessment}. Generally, the color difference is evaluated using the Euclidean distance between two color points in a color space.
As already mentioned, we use the CIELab color space which describes all the
colors visible to the human eye. But in this color space, a small
Euclidean distance between two color points is proportional to the
difference that human visual system perceives. The paper~\cite{ccm} determines
that a larger Euclidean distance has no meaning, only a large difference for the human visual system does. Thus, we employ the color difference proposed in \cite{ccm}.
It is defined as:
\begin{equation}
    D(c(i,j),c(x,y)) = 1 - exp \left(\frac{-E(c(i,j),c(x,y))}{\gamma} \right)
\end{equation}
 where $\gamma$ is the normalized factor, and $E$ is the euclidean distance in CIELab.
\textcolor{black}{ What is the intuition behind this formula?}
 
 \textcolor{black}{First, if we consider identical pixels, then $c(i,j) = c(x,y)$, so Euclidean distance is zero, therefore $D(c(i,j), c(x,y)) = 0$.
 Conversely, if pixels strongly differ, then $exp \left(\frac{-E(c(i,j),c(x,y))}{\gamma} \right) \approx 0$. Then $D(c(i,j), c(x,y)) = 1$.
 Overall, we need the exponent to capture that high Euclidean distance does not correspond with strong difference in human perception.}
 
\textcolor{black}{Overall CCM of image $\mathcal{I}$ is computed as mean CCM over all pixels:
\begin{equation}
    CCM = \frac{1}{\vert \mathcal{I} \vert} \sum_{i,j \in \mathcal{I}} \psi(i, j) 
\end{equation}}

\begin{figure}[h!]
    \centering
    \subfloat[\centering Seated Figure with Architecture Background]{{\includegraphics[width=4cm, height=5cm]{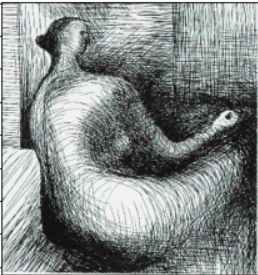}}}%
    \qquad
    \subfloat[\centering Setting sun]{{\includegraphics[ width=4cm, height=5cm]{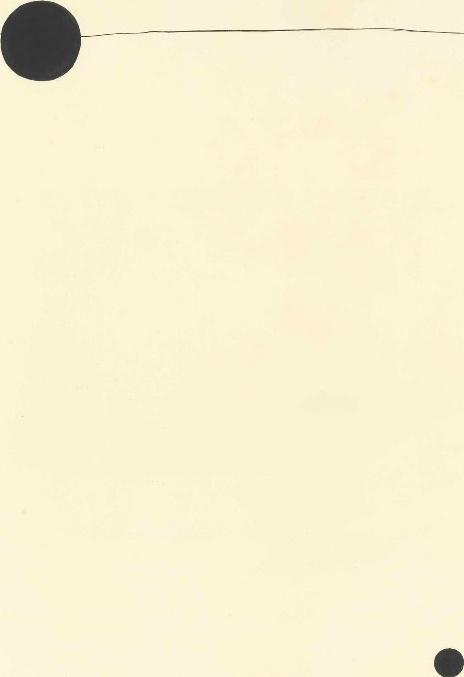}}}%
    \caption{Painting on the left has a high CCM, whereas painting on the right  has a low CCM.}%
    \label{fig:high_low_ccm}%
\end{figure}

\begin{figure}[h!]
    \centering
    \subfloat[\centering CCM=5.7]{{\includegraphics[width=4cm, height=5cm]{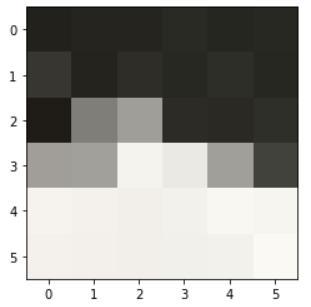}}}%
    \qquad
    \subfloat[\centering  CCM=0.4]{{\includegraphics[ width=4cm, height=5cm]{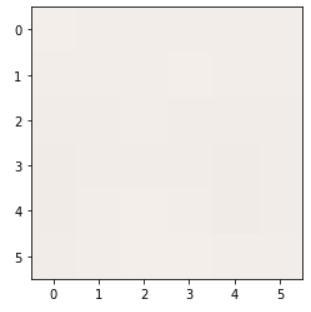}}}%
    \caption{\textcolor{black}{Example of local windows of $6 \times 6$ pixels with corresponding values of CCM. Local window on the left has drastic color change, whereas local window on the right does not show any color change at all, hence CCM of the left painting is higher.} }%
    \label{fig:local_ccm}%
\end{figure}

\textcolor{black}{Examples of CCM are illustrated in Figure \ref{fig:high_low_ccm} and Figure \ref{fig:local_ccm}.
CCM of ``Painting with White Lines'' is higher than CCM of ``Pink in Gray'' (Figure \ref{fig:example_kandinsky}). This result is expected, since Painting with White Lines is clearly harder to paint in terms of visual details.  }

\subsubsection{Points of interest}
 Points of interest are a reflection of the contents of an image.
When many points of interest exist in an image, the complexity of the image is perceived to be high \cite{points}. 

We will measure local symmetry with discrete symmetry transform (DST)\cite{points}.
\textcolor{black}{Local symmetry has shown interesting properties in detecting points of interest to which visual attention may be drawn.
The DST extracts zones of the image in which the local gray levels show a high degree of radial symmetry (where the
degree of locality depends on the radius of the local window). It is worth noting that points of interest detected
by DST appear to be related with points to which the gaze of humans watching the
same image shift~\cite{points}. Apart from the natural attraction of symmetry, this also means that the more the points of interest in an
image, the more complex the image is perceived to be.
More specifically, DST computes local symmetries of an image based on a measure of axial moments of a body
around its center of gravity. In the image case, the pixels inside a circular window are considered as point masses.}
Suppose $r$ is radius and axial moment $n$ with slope $\frac{\pi k}{n}$ are given for $k = 0, \dots, n$. By $g_{x,y}$ we denote a gray level of pixel $(x, y)$.
\begin{equation}
    DST_{i,j} = E(i,j) \times T(i,j)
\end{equation}
Local smoothness of the image is measured as 
\begin{equation}
    E(i,j) = \sum_{(l,m) \in C_r, (p, q) \in C_{r+1}}  \abs{g_{l, m} - g_{p, q}} 
\end{equation}

where $C_r$ is a circle with its center in  $(i,j)$ and radius $r$, moreover, $(l-p)^2 + (m-q)^2 = 1$

\begin{equation}
    T(i, j) = 1 - \sigma(T_k(i,j)),
\end{equation}

where $T_k(i,j)$ is the first order moment relative to an axis with
orientation $\frac{\pi k}{n}$. 

\textcolor{black}{Low standard deviation of moments with different axes imply that moments do not change much depending on a particular axis, which is a sign of symmetry. On the other hand high standard deviation of moments with different axis implies asymmetry:}

\begin{equation}
    T_k(i,j) = \sum_{(l,m) \in C_r}  \abs{(i-l) sin \left( \frac{\pi k}{n} \right) - (j-m)cos \left(\frac{\pi k}{n} \right)} \times g(l,m)
\end{equation}

\begin{figure}
    \centering
    \subfloat{{\includegraphics[scale=0.4]{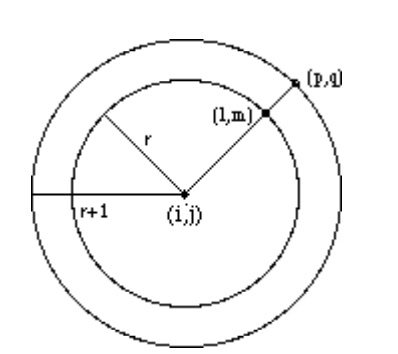}}}%
    \caption{Discrete symmetry transform. Start with center of circle $(i,j)$ and consider two circles with radii $r$ and $r+1$. Next, calculate local smoothness and first order moment to obtain DST. The image was obtained from \cite{points}.}%
    \label{fig:DST}%
\end{figure}

Points of interest are obtained as 
\begin{equation}
p_{i, j} = 
    \begin{cases}
        DST_{i, j} & DST_{i,j} \geq \mu + 3\sigma, \\ 
        0 & otherwise
    \end{cases}\,
\end{equation}

where $\mu$  and $\sigma$ are expectation and standard deviation of $DST$ calculated on the entire painting.

\begin{figure}[h!]
    \centering
    \subfloat[\centering  There are 31 points of interest]{{\includegraphics[ width=5cm, height=5cm]{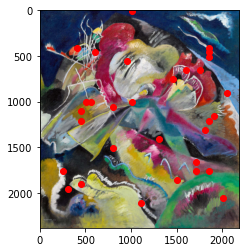}}}%
    \qquad
        \subfloat[\centering There are 3 points of interest]{{\includegraphics[width=5cm, height=5cm]{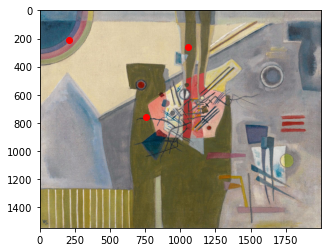}}}%
  \caption{\textcolor{black}{ Expensive and inexpensive artwork of Kandinsky with points of interest colored in red. Paintings on the left has a more diverse color composition than painting on the right, hence the amount of points of interest on the expensive painting is higher.}}%
    \label{fig:DST_paintings}%
\end{figure}

\textcolor{black}{Illustration of DST is given in Figure \ref{fig:DST}. Points of interest of Kandinsky paintings are presented in Figure \ref{fig:DST_paintings}. We see that the painting with a higher price has more points of interest.}

\subsubsection{Edge Density} 
An edge in an image is a significant local change in the image intensity.
An edge image has two types of pixels: edge and non-edge.
Edge density is calculated using the canny edge detection algorithm \cite{canny1986computational}:
\begin{enumerate}
\item Apply a Gaussian filter to smooth the image and remove the noise.
The equation for a Gaussian filter kernel of size $(2k+1) \times (2k+1)$ is given by:
\begin{equation}
     H_{i,j} = \frac{1}{2 \pi \sigma^2}
 \exp\biggl(\,  - \frac{(i-(k+1))^2 + (j-(k+1))^2}{2\sigma^2} \biggr)
\end{equation}

\item  Find the intensity gradients of the image A:

\begin{equation}
     G_x =  \begin{bmatrix}
    1 & 0 & -1 \\
    2 & 0 & -2 \\
    1 & 0 & -1
\end{bmatrix} * A \; \; \; \; \;
G_y =  \begin{bmatrix}
    1 & 2 & 1 \\
    0 & 0 & 0 \\
    -1 & -2 & -1
\end{bmatrix} * A
\end{equation}
\begin{equation}
    G = \sqrt{G_{x}^2 + G_{y}^2}
\end{equation}
 
Gradient direction is given by:
\begin{equation}
    \theta = arctg \biggl(\frac{G_y}{G_x} \biggr)
\end{equation}
\item Apply gradient magnitude thresholding or lower bound cut-off suppression to get rid of spurious response to edge detection.

\item Apply double threshold to determine potential edges.
\item Track edge by hysteresis: Finalize the detection of edges by suppressing all the other edges that are weak and not connected to strong edges. 
\end{enumerate}

\begin{figure}
    \centering
    \subfloat{{\includegraphics[scale = 0.9]{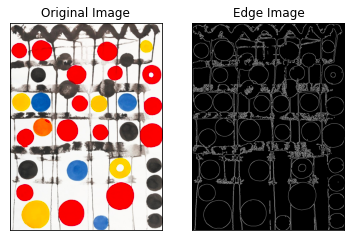}}}%
    \caption{On the left is the original painting, and on the right it has been transformed by the canny edge detection algorithm.}%
    \label{fig:canny}%
\end{figure}

\begin{figure}[h!]
    \centering
    \subfloat[\centering Original image]{{\includegraphics[width=4cm, height=5cm]{Pink_in_Gray.jpg}}}%
    \qquad
    \subfloat[\centering Edge image with density = 0.05]{{\includegraphics[ width=4cm, height=5cm]{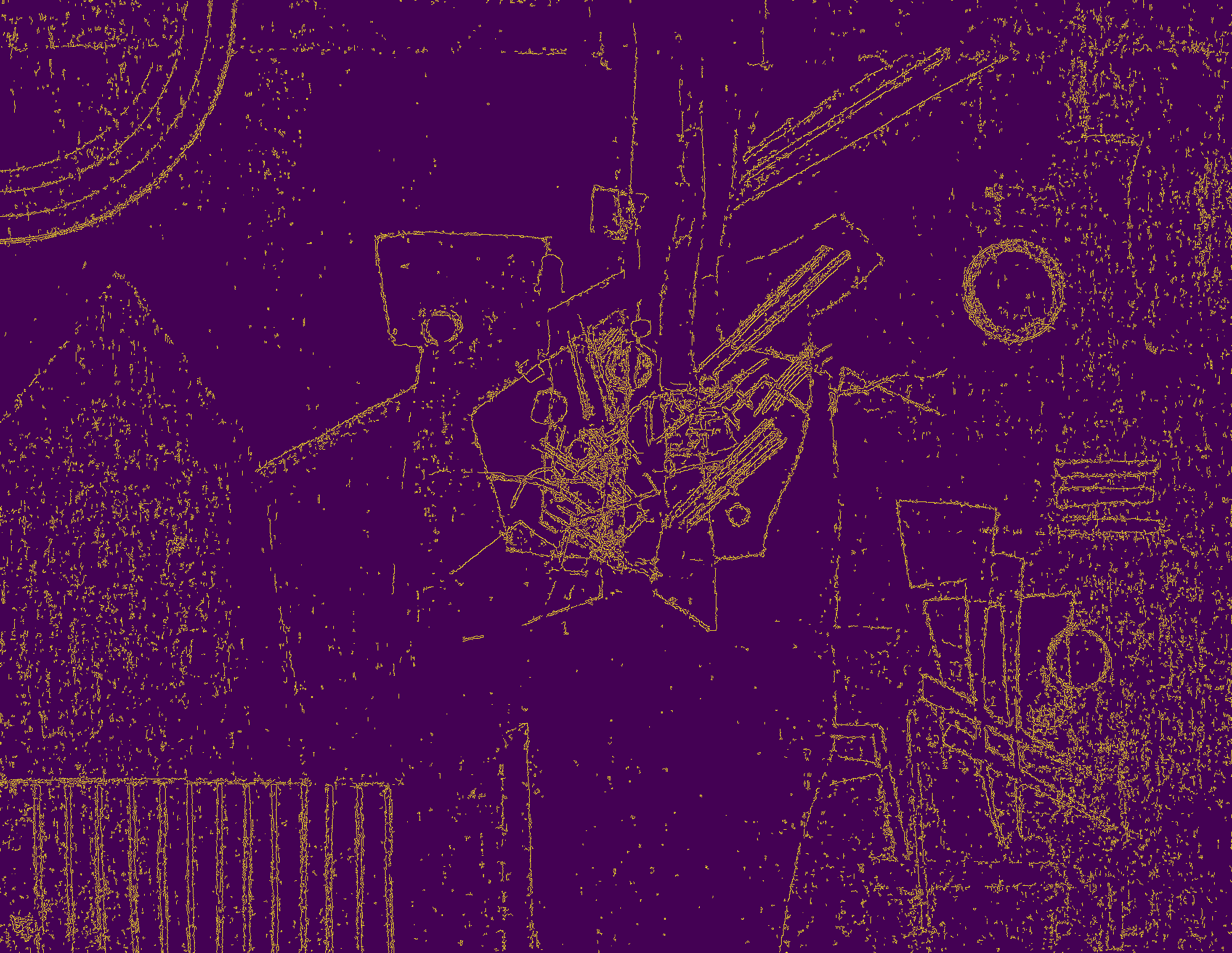}}}%
    \caption{\textcolor{black}{Example of canny edge detection transformation using the inexpensive artwork of Kandinsky. Original artwork has many areas without edges, hence density of edges is relatively small.}}%
    \label{fig:canny_kandinsky}%
\end{figure}

Edge density is calculated as $\frac{N_{edges}}{N_{img}}$, where $N_{edges}$ is the number of edges. 
\textcolor{black}{An example of canny edge detection is shown in Figure \ref{fig:canny}. Edge detection is applied to one of the Kandinsky paintings in Figure \ref{fig:canny_kandinsky}.}

\begin{figure}[h!]
    \centering
    \subfloat[\centering Original image]{{\includegraphics[width=4cm, height=5cm]{Painting_with_White_Lines.jpg}}}%
    \qquad
    \subfloat[\centering Edge image density = 0.12]{{\includegraphics[ width=4cm, height=5cm]{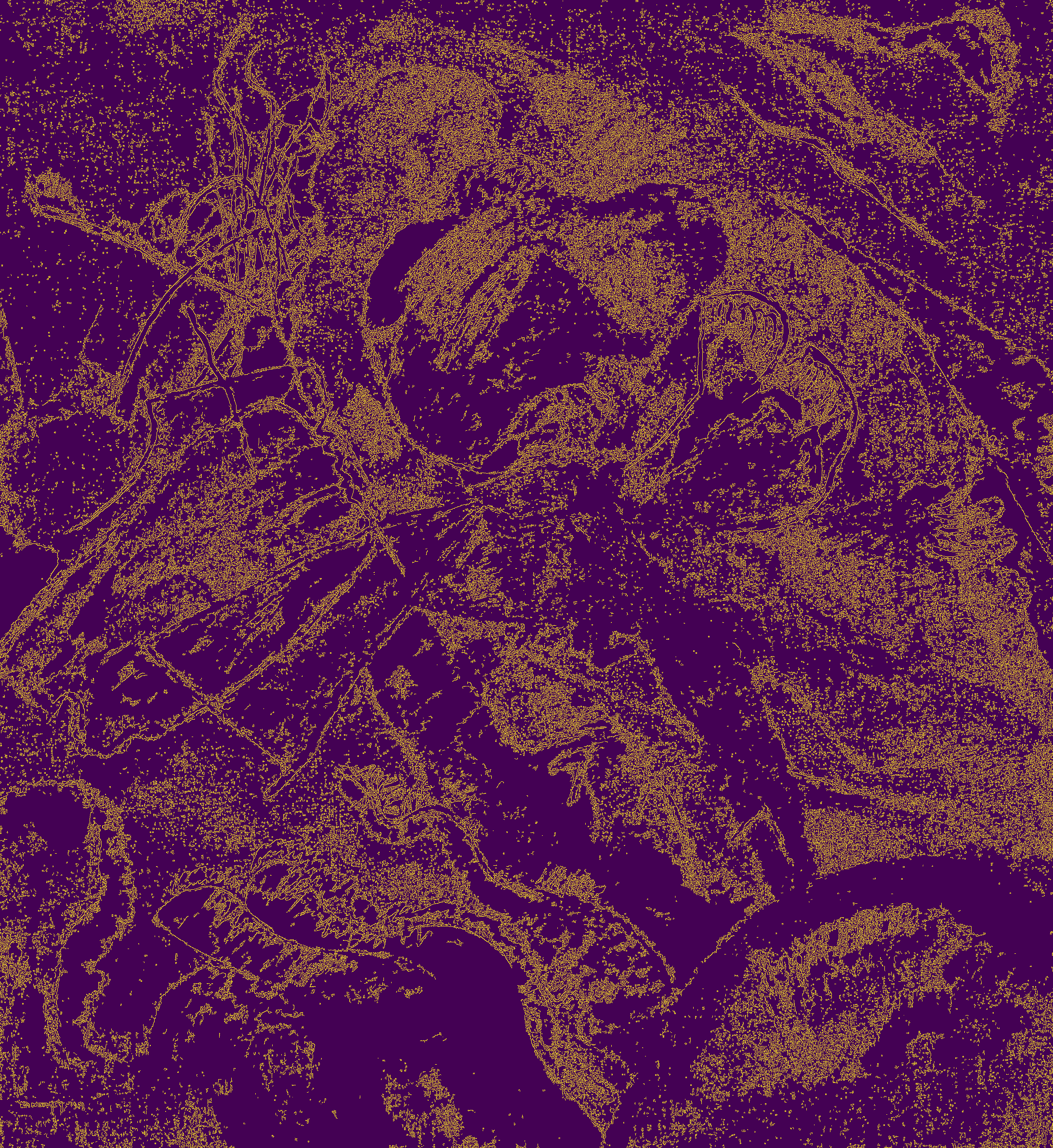}}}%
    \caption{\textcolor{black}{Example of canny edge detection transformation using an expensive artwork of Kandinsky. Original artwork does not have many areas without edges, hence density of edges is relatively high. }}%
    \label{fig:example}%
\end{figure}

\subsubsection{Itten features}
Johannes Itten, a Swiss artist, a theorist of art and an art teacher, gained worldwide fame thanks to the course he created in 1919 in Bauhaus (Staatliche Hochschule für Bau und Gestaltung), which formed the basis of teaching in many modern art schools.

At the beginning of the XX century, Itten's findings on the theory of color had great impact the ideas of artists and scientists about this subject. Itten, based on the experience of his predecessors, both the works of famous artists and even the chemical composition of pigments, developed his own methods and schemes for working with color. He did not claim to be a pioneer in this topic, but instead declared openly that he systematized the works of his predecessors and interpreted them in his own way. Today, the Itten’s sphere is used as a universal tool for working with color. 

In his book ``The Art of Color'', Itten suggested using a twelve-part circle to create color consonances~--- in other words, eye-pleasing color combinations~\cite{itten1961art}. 
All 12 colors are arranged in a circle so that contrasting tones are located in opposite zones according to the rainbow principle~--- from red to purple. Itten's sphere is based on three primary colors – blue, yellow, and red. When mixed in equal proportions, three secondary colors are obtained – purple, orange, and green. After mixing the primary and secondary colors, six colors of the tertiary order are obtained. 

Today, a combination of three colors according to the $60\times10\times30$ principle is considered to be classic. It provides the artworks with an organic and expressive look.

According to theory, painting that has a color distribution similar to a pattern from the color wheel is harmonic from human perspective.
We will extract six features corresponding to six patterns in the color wheel: namely, tetrad, analogue triad, contrast triad, classic triad, complementary, and rectangle patterns. Additionally, we will calculate frequencies of blue, red, yellow and black colors as features.

\begin{figure}[h!]
    \centering
    \subfloat[\centering Original image]{{\includegraphics[width=4cm, height=5cm]{Pink_in_Gray.jpg}}}%
    \qquad
    \subfloat[\centering Frequency decomposition]{{\includegraphics[ width=4cm, height=5cm]{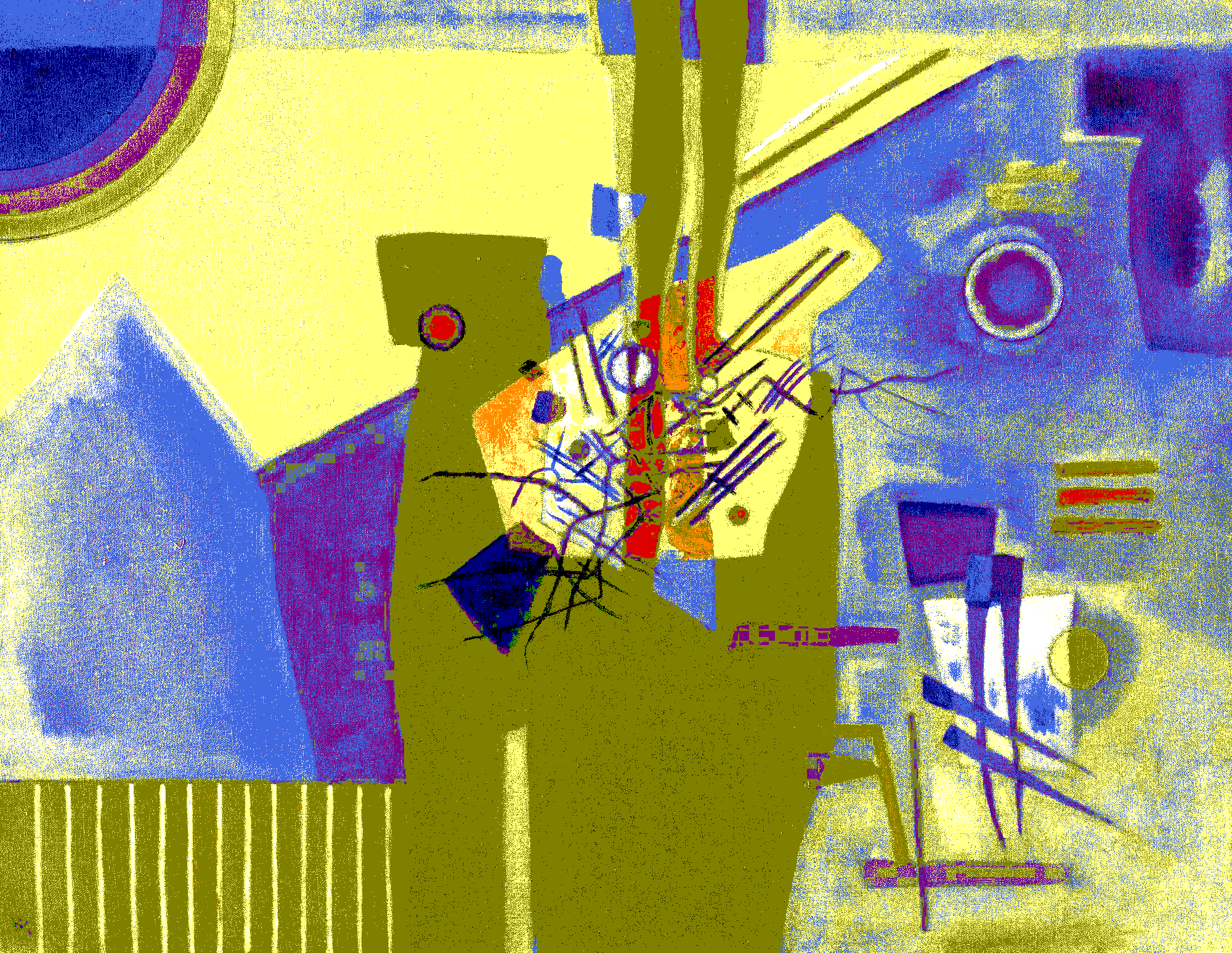}}}%
    \caption{\textcolor{black}{Example of Frequency decomposition of the inexpensive artwork of Kandinsky. Original colors were mapped to the closest ones from Itten color wheel.}}%
    \label{fig:low_itten}%
\end{figure}

\begin{figure}[h!]
    \centering
    \subfloat[\centering Original image]{{\includegraphics[width=4cm, height=5cm]{Painting_with_White_Lines.jpg}}}%
    \qquad
    \subfloat[\centering Frequency decomposition]{{\includegraphics[ width=4cm, height=5cm]{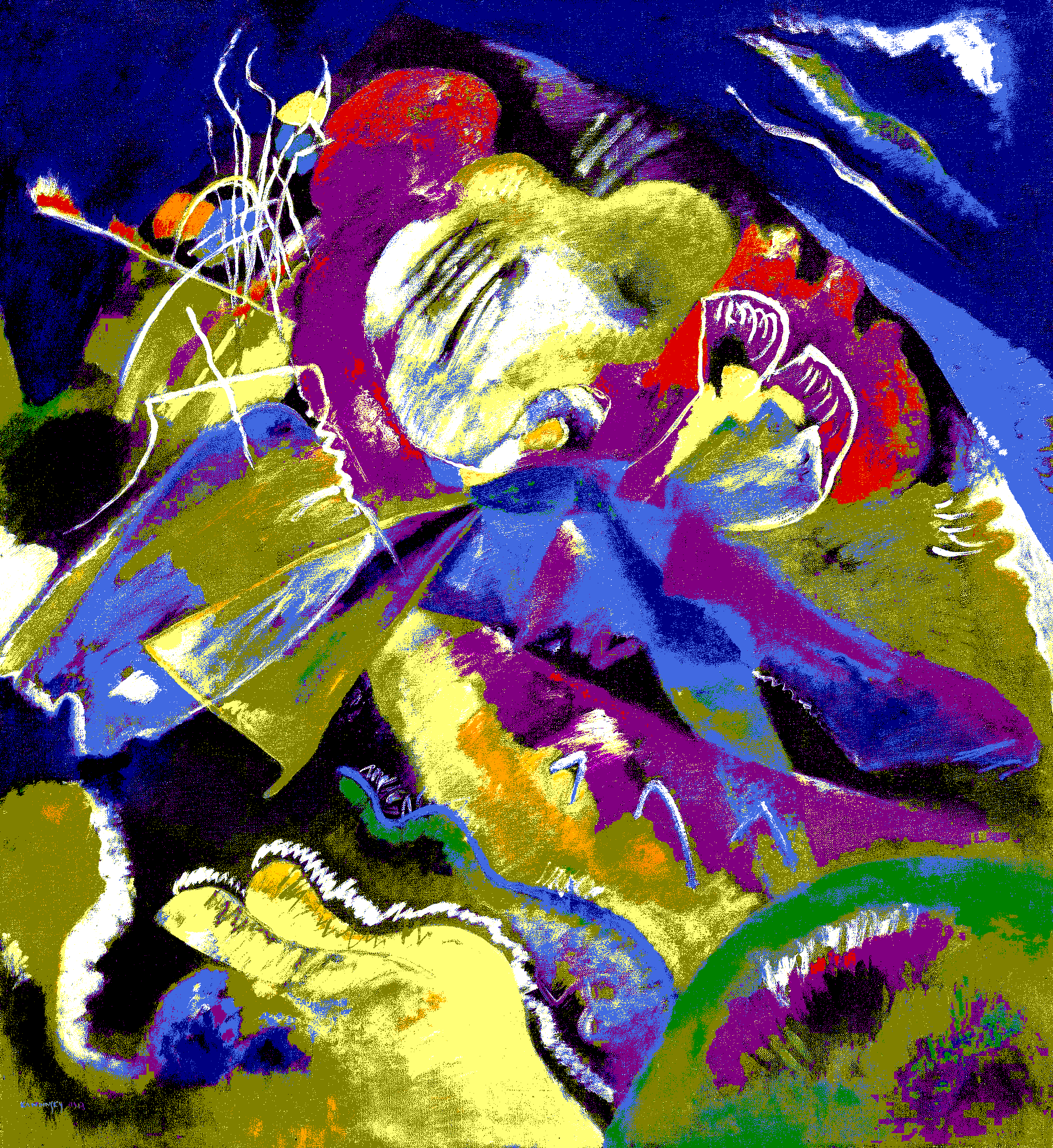}}}%
    \caption{\textcolor{black}{Example of frequency decomposition of a high price painting, using the expensive artwork of Kandinsky. Original colors were mapped to closest ones from Itten color wheel.}}%
    \label{fig:high_itten}%
\end{figure}

In order to assign harmonic patterns to images we firstly decompose them by color frequencies. Then, those images that has high frequencies of colors from a harmonic pattern considered harmonic according to chosen pattern.
\textcolor{black}{Examples of color frequencies decomposition can be seen in Figure \ref{fig:high_itten} and in Figure \ref{fig:low_itten}. Both paintings are found to be not harmonic according to Itten.}
\begin{figure}[h]
\centering
\includegraphics[width=.3\textwidth]{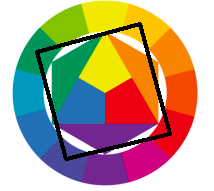}\hfill
\includegraphics[width=.3\textwidth]{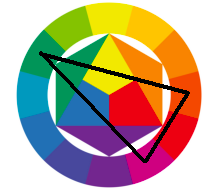}\hfill
\includegraphics[width=.3\textwidth]{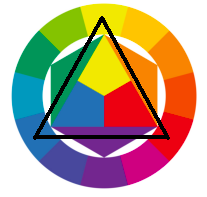}
\caption{Harmonic patterns on the color wheel: tetrad, contrast triad, and classic triad, in which the colors were obtained from the corresponding corners. Any rotation of a square or a triangle is considered to be harmonic.}
\label{fig:figure3}
\end{figure}

\begin{figure}[h]
\centering
\includegraphics[width=.3\textwidth]{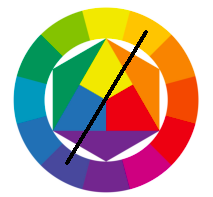}\hfill
\includegraphics[width=.3\textwidth]{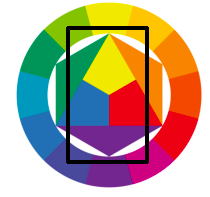}\hfill
\includegraphics[width=.3\textwidth]{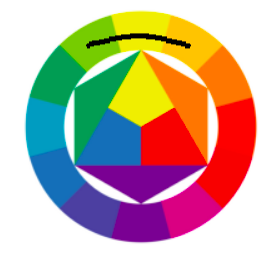}
\caption{Harmonic patterns on the color wheel: complementary, rectangle, and analogue. Any rotation of these shapes is considered to be harmonic.}
\label{fig:figure3}
\end{figure}

\begin{table}[h!]\label{Table 4}
\begin{center}
\begin{tabular}{lrrrrrr}\toprule                       &  Points  & CCM & Lines variance& Harmonic   \\
\midrule 
 Kandinsky expensive               & 31 &    1.2 & 0.12 &      0\\
 Kandinsky inexpensive               & 3 &    1.0 &  0.05 &       0\\ 
 Moore  expensive            &  8 &    1.0 &   0.09 &       0  \\ 
 Moore  inexpensive          &  7 &    2.0 &   0.19 &       0 
 \\\hline\end{tabular}
 \caption{\textcolor{black}{Examination of indexes on Kandinsky and Moore expensive and inexpensive artworks. Generally, Kandinsky paintings are worth more then Moore paintings. }}

\end{center}
\end{table}

\begin{figure}[h!]
    \centering
    \subfloat[\centering Two women and child, 1948,  \$2.96 million]{{\includegraphics[width=4cm, height=5cm]{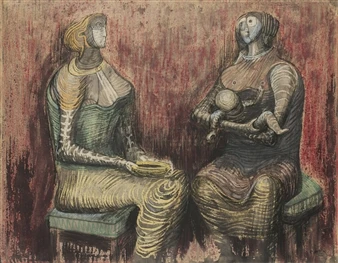}}}%
    \qquad
    \subfloat[\centering  Shelter Drawing: Seated Mother and Child, 1942,     \$0.86 million]{{\includegraphics[ width=4cm, height=5cm]{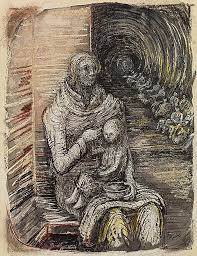}}}%
    \caption{\textcolor{black}{An expensive and an inexpensive artworks of Moore. Painting on the right is more complex visually than the painting on the left, hence its color complexity measure is higher.} }%
    \label{fig:moore}%
\end{figure}

\subsection{Segmentation} 
To obtain local features, we need to segment paintings. We employ Felzenszwalb segmentation as initial segmentation~\cite{felzenszwalb2004efficient}.
Next, we perform the Vertices-Graph Growing Process to produce more accurate segments.
\subsubsection{Initial Segmentation}

Let $G = (V, E)$ be a graph, where $V$ are the pixels, and $E$ is the measure of similarity between pixels. A segmentation $S$
is a partition of $V$ into components such that each component (or region) $C \in S$ corresponds to a connected component in a graph $G$. 

Internal difference of a component
$C \subset  V$ is defined as  the largest weight in the minimum spanning tree of the component, $MST(C, E)$. That is,
\begin{equation}
    Int(C) = \max_{e \in MST(C,E) } w(e)
\end{equation}
 
 Next, the difference between two components $C_1,C_2 \subset V$ is defined as the edge of minimum weight that connects them.
\begin{equation}
     MInt(C_{1}, C_{2}) = min(Int(C_1) + \tau(C_1),Int(C_2) + \tau(C_2) )
\end{equation}
where $\tau = \frac{k}{\abs{C}}$, and $k$ is some constant parameter. 

\begin{algorithm}
\caption{Segmentation}
\begin{algorithmic}[1]
\Require Graph $G = (V, E)$, with $n$ vertices and $m$ edges
\Ensure Segmentation of V into components $S = (C_1,...,C_r)$
\State Sort $E$ into $\pi = (o_1,\dots, o_m)$, by non-decreasing
edge weight.
\State Start with a segmentation $S_0$, where each vertex $v_i$
is in its own component.

\While{$q = 1..m$}
        \State Let $o_q = (v_i, v_j)$
         \State $C^{q-1}_i$ is component of $S^{q-1}$, containing $v_i$ 
        \State  $C^{q-1}_j$ is component of  $S^{q-1}$ containing $v_j$  
        
         \If{$C^{q-1}_i \neq C^{q-1}_j$  AND  $w(o_q) \leq MInt(C^{q-1}_i ,C^{q-1}_j )$}
            \State $S^q = C^{q-1}_i \cup C^{q-1}_j $
        \Else
            \State  $S^q = S^{q-1}$
        \EndIf
\EndWhile
\end{algorithmic}
\end{algorithm}

After the initial segmentation, we merge components using the Regions adjacency graph~\cite{tremeau2000regions}.

\subsubsection{Regions adjacency graph}
Let $G = (V,E)$, where $V$ is a set of regions and $E$ is the set of edges between them. 
Fisher distance can be used for measuring similarity between regions~\cite{schettini1993segmentation}:


\begin{equation}
    FD_{kij} = \frac{\sqrt{n_i + n_j}\abs{\mu_{ki} - \mu_{kj}}}{\sqrt{n_i \sigma^2_{ki} + n_j \sigma^2_{kj}}}, \quad \text{if} \; \sigma^2_{ki}, \sigma^2_{kj} \neq 0, 
\end{equation}
\begin{equation}
    FD_{kij} =\abs{\mu_{ki} - \mu_{kj}}, \quad \text{if}  \; \sigma^2_{ki} = \sigma^2_{kj} = 0
\end{equation}
 where $n_k$ is size of region $h$, and $\mu_{kh}$ and $\sigma^2_{kh}$ are expectation and variance of feature $k$ in region $h$. Let
 \begin{equation}
     FD_{ij} = \max_{k}(C_k FD_{kij})
 \end{equation}
  where $C_k$ is either zero or one. Good color separation is obtained when $FD_{ij} \geq 4$.

\subsubsection{Vertices-Graph Growing Process}
A vertice $v_i$ could be merged with one of its
neighbors $v_j \in V$ if
\begin{enumerate}
    \item $d^2(R_i; R_j )$  is sufficiently small;
    \item $d^2(R_i; V)$ is sufficiently small, $V$ are neighbors of $v_i$;
    \item $d^2(R_i; S)$ is sufficiently small, $S$ is the subset of $V$ which contains $v_j$;
\end{enumerate}
 where $d^2(R_i; R_j ) = FD_{ij}$.

\begin{figure}[H]
    \centering
    \subfloat{{\includegraphics[width = 12cm, height = 5 cm]{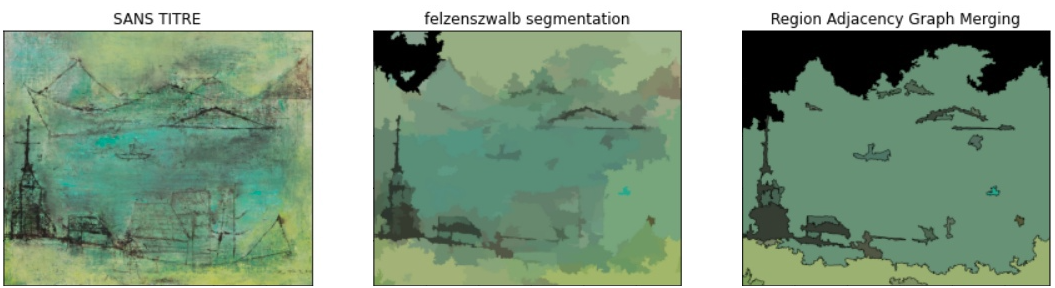}}}%
    \caption{An example of the segmentation process, shown on the \textit{Sans titre} painting by Albert Féraud. The leftmost panel contains the original image, the middle panel is the image transformed with Felzenszwalb segmentation, and in the rightmost panel, the Regions adjacency graph merging is applied. }%
    \label{fig:example}%
\end{figure}

\subsection{Local color features}

Color features are important not only for the global impression but
 also for the local details. In this part, the local color features are represented through the average of hue, saturation and lightness for the
 two largest segments and the color contrast between the largest segments and their neighbor segments.
 
Hue, saturation and lightness of the two largest segments are obtained as:
\begin{equation}
    h_{fls} = \frac{1}{AREA_{fls}}\sum_{(m,n) \in AREA_{fls}} h(m,n)
\end{equation}
\begin{equation}
    s_{fls} = \frac{1}{AREA_{fls}}\sum_{(m,n) \in AREA_{fls}} s(m,n)
\end{equation}
\begin{equation}
    v_{fls} = \frac{1}{AREA_{fls}}\sum_{(m,n) \in AREA_{fls}} v(m,n)
\end{equation}
where $AREA_{fls}$ is a number of pixels in the first largest segment, $h(m,n)$, $s(m,n)$, $v(m,n)$ are hue, saturation and lightness of pixel $(m,n)$.

We will measure painting contrast as maximum distance from hue (saturation or value) on the largest segment to corresponding coordinate on the neighbour segments:
\begin{equation}
    contrast_h = max \abs{ H_{largest} - H_i }, i \in \Omega_{neib}
\end{equation}
\begin{equation}
    contrast_s= max \abs{S_{largest} - S_i}, i \in \Omega_{neib}
\end{equation}
\begin{equation}
    contrast_v = max \abs{V_{largest} - V_i}, i \in \Omega_{neib}
\end{equation}

Shape complexity of the first and second largest segments are obtained as 
\begin{equation}
   c_i =  \frac{P_{R_i}^2}{4 \pi A_{R_i}}
\end{equation}
where $P_{R_i}$ and $A_{R_i}$  are the perimeters and areas of the first and second largest segments. Additionally, we include areas of the first and second largest segments and number of segments in segmentation as features.

\section{Empirical results}\label{sec6}
\begin{figure}
    \centering
    \includegraphics[width=10cm, height=12cm]{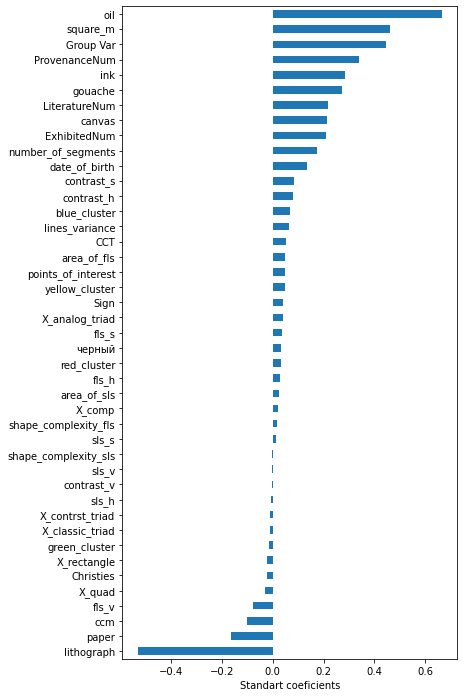}
    \caption{Bar plot of standard coefficients from linear mixed model. }
    \label{fig:coef_standart}
\end{figure}

Let us now consider the results of our linear mixed  model. First, we examine the results of classic price determinants and then we study how the differences 
in art prices across paintings are influenced by color. \hyperref[Table 5]{Table 5} presents the numerical results. Figure \ref{fig:coef_standart} presents bar plot of standardized coefficients from linear mixed model. We will judge significance by p-values and standardized coefficients.
\textcolor{black}{Let us consider Kandindsky and Moore inexpensive and expensive paintings \hyperref[Table 4]{Table 4}. Moore paintings are shown on Figure \ref{fig:moore}. We considered Moore as Author with inexpensive paintings in our data. First of all, all four paintings are not harmonic according to Itten color theory. It may give some sense towards its irrelevance. Secondly, points of interest on expensive paintings are higher then on inexpensive ones. Next, CCM is highest on Moore's inexpensive paintings. It may indicate that it was painted with plenty of details. Lastly, lines variance is highest on the inexpensive Moore painting for the same reason that it has high CCM. Overall, the results from these examples confirm our findings using statistical modelling. For example, higher CCM is in fact associated with a lower price.}

A painting's size, its number of exhibitions, mentions in the literature, owners and the presence of a signature are all positively associated with the price. Images of abstract paintings were obtained from either Sotheby's or Christie's. Origin is found to be not significant, as well as date of author's birth. Among used materials, namely, canvas, oil, paper, ink, gouache, lithograph, only paper and lithograph are inferior in price to the others.

Abstract paintings that used the analogue triad harmonic pattern are superior in price. However, other harmonic patterns are found to be insignificant. Frequency of the blue color is carrying premium, whereas red, green, yellow and black are not. Multiple metrics of color complexity are found to be not significant, e.g. points of interest and lines variance. For first largest segment they are area, hue, saturation, shape complexity. For second largest segment they are area, hue, saturation, value, shape complexity. Negatively significant color complexity metrics are color complexity measure and value on first largest segment. 

Positively significant color diversity metrics are
contrast in hue, contrast in saturation and number of segments, with the exception of contrast in value which is insignificant. Finally, correlated color temperature has low explanatory power for abstract paintings.

\begin{table}[h!]\label{Table 5}
\caption{Mixed Linear Model Regression Results}
\begin{center}
\begin{tabular}{lrrrrrr}\toprule                       &  Coef. & Std.Err. &       z & P $>$ |z| & [0.025 & 0.975]  \\
\midrule 
Intercept              & 12.044 &    0.206 &  58.338 &       $0.000^{***}$ & 11.639 & 12.449  \\ Christies              & -0.023 &    0.020 &  -1.201 &       0.230 & -0.062 &  0.015  \\ ExhibitedNum           &  0.225 &    0.024 &   9.537 &       $0.000^{***}$ &  0.179 &  0.271  \\ LiteratureNum          &  0.233 &    0.024 &   9.571 &       $0.000^{***}$ &  0.185 &  0.281  \\ ProvenanceNum          &  0.340 &    0.022 &  15.774 &       $0.000^{***}$ &  0.298 &  0.383  \\ Sign                   &  0.044 &    0.019 &   2.271 &       $0.023^{*}$ &  0.006 &  0.082  \\ X\_analog\_triad       &  0.068 &    0.030 &   2.293 &       $0.022^{*}$ &  0.010 &  0.126  \\ X\_classic\_triad      & -0.017 &    0.027 &  -0.628 &       0.530 & -0.069 &  0.035  \\ X\_comp                &  0.031 &    0.029 &   1.077 &       0.281 & -0.026 &  0.088  \\ X\_contrst\_triad      & -0.012 &    0.028 &  -0.422 &       0.673 & -0.066 &  0.043  \\ X\_quad                & -0.045 &    0.025 &  -1.790 &       0.073 & -0.095 &  0.004  \\ X\_rectangle           & -0.034 &    0.027 &  -1.258 &       0.209 & -0.088 &  0.019  \\ area\_of\_fls          &  0.047 &    0.028 &   1.669 &       0.095 & -0.008 &  0.101  \\ area\_of\_sls          &  0.026 &    0.022 &   1.196 &       0.232 & -0.017 &  0.068  \\ blue\_cluster          &  0.227 &    0.057 &   3.957 &       $0.000^{***}$ &  0.115 &  0.340  \\ canvas                 &  0.216 &    0.030 &   7.146 &       $0.000^{***}$ &  0.157 &  0.275  \\ ccm                    & -0.102 &    0.034 &  -2.985 &       $0.003^{***}$ & -0.169 & -0.035  \\ contrast\_h            &  0.079 &    0.024 &   3.315 &       $0.001^{***}$ &  0.032 &  0.125  \\ contrast\_s            &  0.085 &    0.026 &   3.261 &       $0.001^{***}$ &  0.034 &  0.136  \\ contrast\_v            & -0.005 &    0.025 &  -0.183 &       0.855 & -0.053 &  0.044  \\ date$\_$of$\_$birth        &  0.134 &    0.200 &   0.667 &       0.505 & -0.259 &  0.526  \\ fls$\_$h                 &  0.038 &    0.023 &   1.664 &       0.096 & -0.007 &  0.082  \\ fls\_s                 &  0.039 &    0.025 &   1.577 &       0.115 & -0.010 &  0.088  \\ fls\_v                 & -0.081 &    0.025 &  -3.251 &       $0.001^{***}$ & -0.130 & -0.032  \\ gouache                &  0.267 &    0.028 &   9.616 &       $0.000^{***}$ &  0.212 &  0.321  \\ green\_cluster         & -0.041 &    0.080 &  -0.510 &       0.610 & -0.197 &  0.115  \\ ink                    &  0.286 &    0.026 &  11.093 &       $0.000^{***}$&  0.235 &  0.336  \\ lines\_variance        &  0.063 &    0.038 &   1.663 &       0.096 & -0.011 &  0.138  \\ lithograph             & -0.515 &    0.023 & -22.263 &       $0.000^{***}$ & -0.561 & -0.470  \\ number\_of\_segments   &  0.176 &    0.027 &   6.427 &       $0.000^{***}$ &  0.122 &  0.229  \\ oil                    &  0.669 &    0.034 &  19.702 &       $0.000^{***}$ &  0.602 &  0.735  \\ paper                  & -0.165 &    0.030 &  -5.461 &       $0.000^{***}$ & -0.224 & -0.106  \\ points\_of\_interest   &  0.042 &    0.026 &   1.616 &       0.106 & -0.009 &  0.094  \\ red\_cluster           &  0.053 &    0.072 &   0.740 &       0.459 & -0.087 &  0.194  \\ shape\_complexity\_fls &  0.017 &    0.018 &   0.965 &       0.334 & -0.018 &  0.052  \\ shape\_complexity\_sls & -0.003 &    0.020 &  -0.162 &       0.871 & -0.042 &  0.036  \\ sls\_h                 & -0.004 &    0.022 &  -0.203 &       0.839 & -0.047 &  0.038  \\ sls\_s     &  0.013 &    0.021 &   0.627 &       0.531 & -0.029 &  0.056  \\ sls\_v                 & -0.005 &    0.019 &  -0.282 &       0.778 & -0.043 &  0.032  \\ square\_m              &  0.472 &    0.024 &  19.988 &       $0.000^{***}$ &  0.426 &  0.518  \\ yellow\_cluster        &  0.088 &    0.066 &   1.328 &       0.184 & -0.042 &  0.217  \\ black                 &  0.030 &    0.021 &   1.424 &       0.155 & -0.011 &  0.072 \\ 
CCT           &         0.051   &  0.028  &  1.833  & 0.067 &  -0.004  & 0.105
\\ Group Var              &  0.519 &    0.210 &         &             &        &         \\\hline\end{tabular}
\footnotetext{Levels of significance: * (0.05), ** (0.01), ***(0)}
\end{center}
\end{table}

\section{Discussion and Conclusions}\label{sec7}
The present study was designed to determine the effect of visual characteristics on abstract painting price. This section summarizes our findings and contributions. 

One interesting finding is significance of signature on an abstract painting. When comparing our results to those of older studies, it must be pointed out that presence of signature has unstable explanatory power  \cite{cinefra2019determinants}.

This study supports evidence from previous observations about  insignificance of auction house \cite{pownall2016pricing}. Although, certain limitations of this study could be addressed, such as moderate sample and focus on abstract paintings.

Although the present results clearly support insignificance of the painter's birth date similar to~\cite{cinefra2019determinants}, it is appropriate to recognize this limitation. That is, in our data, this feature ranges from 1866 to 1928, with the median being 1899, so its insignificance could be caused by small variation. 

A strong relationship between size of a painting and its price has been reported in the literature \cite{worthington2004art, stepanova2019impact, cinefra2019determinants}. Our results are in agreement with the reported findings.

Further on, reviewing the literature, we have found data supporting the correlation between the number of exhibitions of a painting, the number of mentions in the literature, the number of owners and the price~\cite{cinefra2019determinants}. Our findings highlight a strong positive significance of these predictors, which represent the cultural importance of a painting.

The current study found clear support for the significance of canvas. This finding broadly supports the work of other researchers in this area \cite{ju2019art, cinefra2019determinants, stepanova2019impact}. Lithography and paper are found to be inferior in terms of price to others, since lithography is a printing process, so copies of such artwork could  be easily produced, and paper demands special preservation conditions, as such paintings are fragile. In addition, sketches are usually depicted on the paper, while the completed idea is transferred by the artist to the canvas. Oil on canvas is the most costly technique that requires special efforts and skill from the artist. There are no two absolutely identical canvases, even if they are author's copies.

Our findings on color distribution at least hint that in abstract paintings blue color might be superior to others. In case of Picasso, similar result was obtained by Stepanova~\cite{stepanova2019impact}. The main limitation is the lack of paintings to make a stronger conclusion.

Overall findings about color harmony are in accordance with previous research \cite{charlin2021general}. That is, color harmony has low significance. However, our results go beyond previous reports, showing that color harmony can hold explanatory power, e.g. analogue triad.

From the results, it is clear that color complexity metrics has low explanatory power. For example, lines variance and points of interest are found to be insignificant. Moreover, higher color complexity measure is associated with lesser price. That is, painting has too complicated color distribution, therefore it becomes less attractive for a potential buyer.

The present study has confirmed the findings about diversity of color composition, that is, contrast in hue, contrast in saturation, and number of segments are positively significant~\cite{stepanova2019impact, charlin2021general}. On the other hand, contrast in value represents color diversity poorly. This may be the reason why it was not found to be significant.


\section*{Statements and Declarations}
\begin{itemize}
    \item The authors declare that they have no conflict of interest
\end{itemize}
\bibliography{sn-bibliography}

\newpage 

\section*{Appendix}
\begin{figure}[h!]
    \centering
    \subfloat{{\includegraphics[width=5cm, height=6cm]{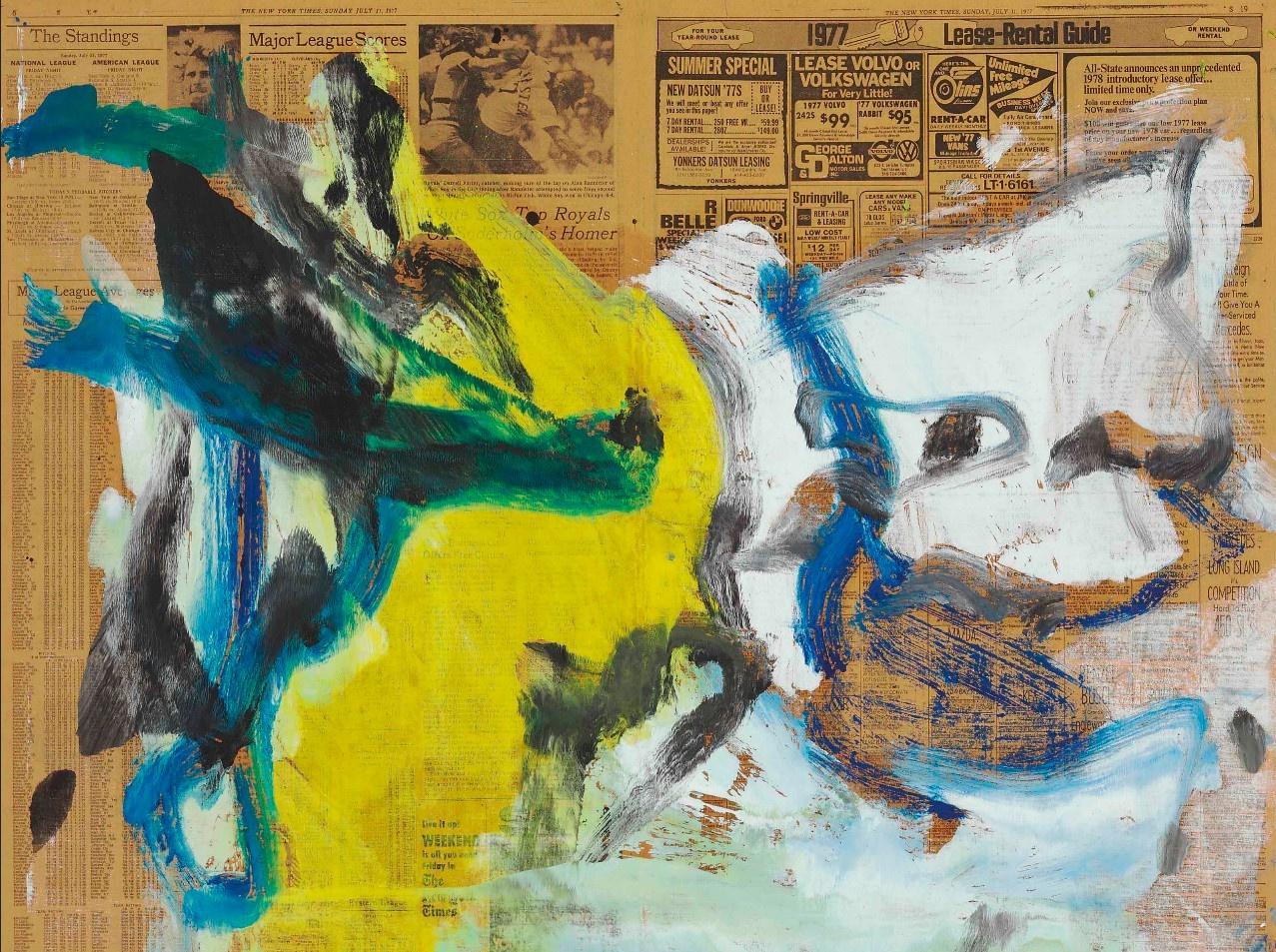}}}%
    \qquad
    \subfloat{{\includegraphics[ width=5cm, height=6cm]{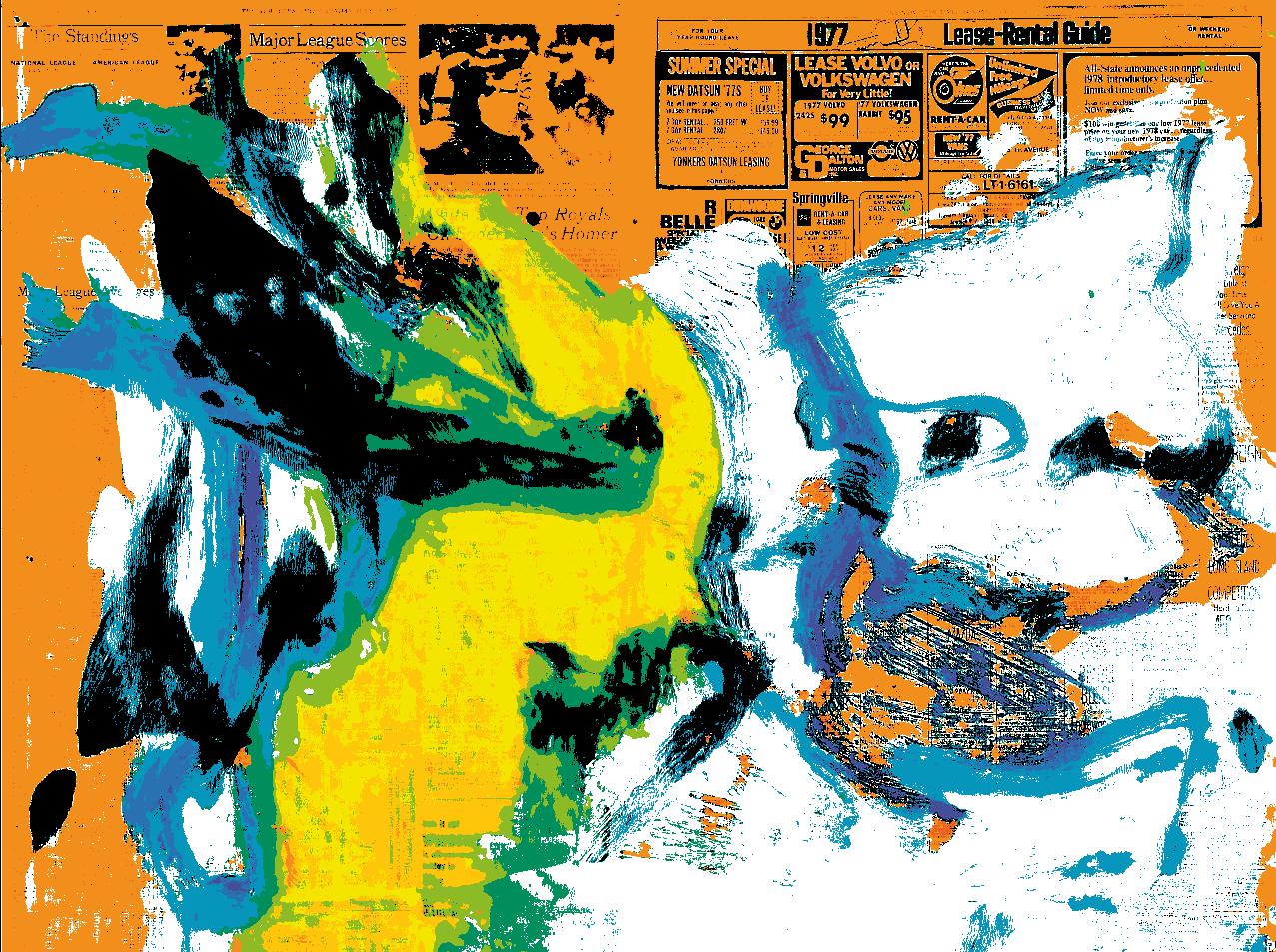}}}%
    
    \vspace*{+2mm}
    \caption{The left is the original painting; the right was obtained by frequency decomposition.}%
    \label{fig:example}%
\end{figure}

\end{document}